\documentclass[aip,sd,amsmath,amssymb,reprint]{revtex4-1}

\usepackage{graphicx}
\usepackage{bm}
\usepackage{color}
\usepackage{ulem}

\begin{document}


\title{Effect of one-dimensional superlattice potentials on the band gap of two-dimensional materials}
\author{Shota Ono}
\email{shota\_o@gifu-u.ac.jp}
\affiliation{Department of Electrical, Electronic and Computer Engineering, Gifu University, Gifu 501-1193, Japan}

\begin{abstract}
Using the tight-binding approach, we analyze the effect of a one-dimensional superlattice (1DSL) potential on the electronic structure of black phosphorene and transition metal dichalcogenides. We observe that the 1DSL potential results in a decrease of the energy band gap of the two-dimensional (2D) materials. An analytical model is presented to relate the decrease in the direct-band gap to the different orbital characters between the valence band top and conduction band bottom of the 2D materials. The direct-to-indirect gap transition, which occurs under a 1DSL potential with an unequal barrier width, is also discussed.
\end{abstract}

\maketitle


\section{Introduction}
Recently, a wide variety of graphene sister materials, such as black phosphorene (BP) \cite{Hliu} and monolayer transition metal dichalcogenides $MX_2$ ($M=$Mo, W; $X=$S, Se, Te) \cite{Radi,QHwang}, have been synthesized. Furthermore, a recent search on the Materials Project database yielded more than 600 stable two-dimensional (2D) materials that can be synthesized by exfoliation.\cite{ashton} Since some of these 2D materials possess a finite band gap, they are expected to be used in a host of next-generation electronic and optoelectronic devices. 

Strain engineering has attracted a lot of attention in the context of providing a tunable band gap of 2D materials.\cite{peng,rodin,jiang,rostami,pearce} For example, the early studies, based on the density-functional theory, have shown that the strain can yield a semiconductor-metal transition in BP.\cite{peng,rodin} An analytical study has shown that the band gap increases (decreases) by a few hundred meV when the tensile (compressive) strain lies in the 2D plane.\cite{jiang} 

In view of the tight-binding (TB) description, the strain application can be interpreted as the hopping parameter modification. How the locally modified on-site potential parameter influences the magnitude of the band gap in 2D materials should then be determined. Thus far, monolayer graphene in a one-dimensional superlattice (1DSL) potential has been extensively investigated based on the Dirac-type \cite{cbai,park1,park2,brey,barbier,wang,burset,maksimova,shakouri,ono,choi,chen,kim} and TB Hamiltonian.\cite{pal,meux} The presence of a 1DSL potential has been shown to be able to drastically change the energy band structure around the Dirac cone, yielding electron supercollimation, extra Dirac cones, and an opening of the band gap. As the graphene superlattices have been fabricated by several experiments, where the period of the superlattice structure is from a few nm up to several hundreds of nm,\cite{dubey,bai,lin,drienovsky} a theoretical study for the 1DSL potential effect on the band gap of 2D materials would be important for future applications. 

Herein, we study the electronic structure of the BP and $MX_2$ in 1DSL potentials. By using the TB models developed by Rudenko and Katsnelson \cite{rudenko} and Liu {\it et al},\cite{liu} we show that the band gap of the 2D materials is lowered by the 1DSL potential. Through a simple model analysis, the decrease in the direct band gap is shown to be attributed to the different orbital characters between the valence band (VB) top and conduction band (CB) bottom. We also show the direct-to-indirect gap transition that occurs under a 1DSL potential with an unequal barrier width. The models examined below can be applied to situations in which the 1DSL potential is controlled by two (top and back) gates,\cite{dubey,drienovsky} or in which the 2D layer is placed on a substrate with an appropriate lattice mismatch.\cite{bai,lin} 

\begin{figure*}[t]
\center
\includegraphics[scale=0.375,clip]{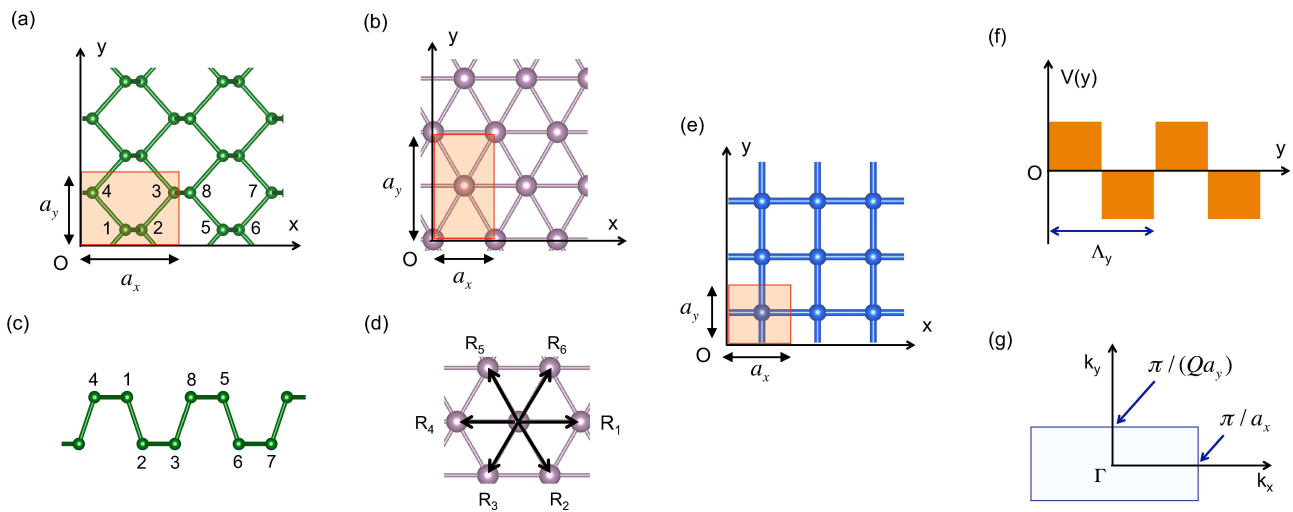}
\caption{\label{fig:network} Schematic illustration of (a) BP and (b) $MX_2$. The conventional unit cells are depicted by the shaded rectangle with the size of $a_x \times a_y$. (c) Side view on the BP structure. The atoms numbered 1, 4, 5, and 8 deviate from the plane at which the atoms numbered 2, 3, 6, and 7 are located. (d) The translation vectors $\bm{R}_i$ with $i=1,2,3,4,5$, and 6 for $MX_2$. (e) Schematic illustration of the 2D square net model. (f) The potential profile along the $y$-direction in the 2D material superlattice. (g) The first Brillouin zone of the 2D material superlattice with a size of $S_Q$ (see text for the definitions).}
\end{figure*}

\section{2D Material Superlattice}
\label{sec:2DMS}
To study the 1DSL potential effect on the electronic band structure of BP and $MX_2$, we first consider a conventional unit cell that consists of $N$ atoms. The area of the unit cell is defined as $S_1\in [0,a_x)\otimes[0,a_y)$, as shown in Fig.~\ref{fig:network}(a) and (b) for BP and $MX_2$, respectively. For example, $a_x=4.37$ and $a_y=3.32$ for BP and $a_x=$3.19 and $a_y=\sqrt{3}a_x$ for MoS$_2$ in units of \AA. Next, we create a 1$\times Q$ supercell, that is, $S_Q \in [0,a_x)\otimes[0,Qa_y)$ with a positive integer $Q$. The position of each atom can be defined as $\bm{R}^{s} = (R_{x}^{s},R_{y}^{s})$ with $s= 1,\cdots, NQ$. The 1DSL potential that changes periodically along the $y$-direction is applied to the 2D materials as 
\begin{equation}
 v_{s} = 
 \begin{cases}
  V_0 & {\rm for} \ \ 0\le R^{s}_{y} < \dfrac{\Lambda_y}{2}  \\
  -V_0 & {\rm for} \ \ \dfrac{\Lambda_y}{2}  \le R^{s}_{y} < \Lambda_y,
 \end{cases}
 \label{eq:SLpot}
\end{equation}
where $\Lambda_y = Qa_y$ [see Fig.~\ref{fig:network}(f)]. In the following, we set $Q=20$, unless noted otherwise. The use of a larger $Q$ (and $\Lambda_y$) does not change the $V_0$-dependence of the band gap size, as demonstrated below. The first Brillouin zone (BZ) has a rectangular shape surrounded by four lines $k_x = \pm \pi / a_x$ and $k_y = \pm \pi / (Qa_y)$, as shown in Fig.~\ref{fig:network}(g). For the pristine BP and $MX_2$ (i.e., the case of $V_0 = 0$ and $Q=1$), the minima of the band gaps (i.e., both the CB bottom and VB top) are located at the $\Gamma$ point and ($2\pi/(3a_x), 0$) in the first BZ, respectively. Since these are located along the $k_y=0$ line, the application of the 1DSL potential, given by Eq.~(\ref{eq:SLpot}) with $Q\ne 1$, does not change the location of the band gap minimum even if zone foldings occur. Below, we will consider only the electronic band structure along the $k_x$-direction.

For the case of the 1DSL potential that changes periodically along the $x$-direction, the location of the band gap minimum is also determined with consideration of the zone-folding concept. Similar to the case above, the band gap minimum is still located at the $\Gamma$ point for the BP superlattices. In contrast, the location of the band gap minimum moves along the $k_x$-direction with $Q$ for the $MX_2$ superlattices. By using the zone-folding concept, one can observe that the band gap minimum is located at the $\Gamma$ point when $Q=3i$ and $(\pm 2\pi/(3Qa_x), 0)$ when $Q\ne3i$ with a positive integer $i$. The most important observation is that the direction of the 1DSL potential does not alter the main result in this work, that is, the band gap reduction arising from the 1DSL potential.

It should be noted that when the period of the 1DSL potential, given by Eq.~(\ref{eq:SLpot}), is large ($Q\gg 1$), the group velocity along the $k_y$-direction is negligibly small compared with that along the $k_x$-direction. This is true when we consider the 1DSL potential that changes periodically along the $x$-direction; given a large period along the $x$-direction, the group velocity along the $k_x$-direction is, in turn, negligibly small compared with that along the $k_y$-direction.

For the study of the $MX_2$ superlattices below, we assume that the 1DSL potential is independent of the orbitals $d_{z^2}$, $d_{xy}$, and $d_{x^2-y^2}$, for simplicity. The application of the orbital-dependent 1DSL potential does not change the main result of this work.

\begin{figure}[t]
\center
\includegraphics[scale=0.45,clip]{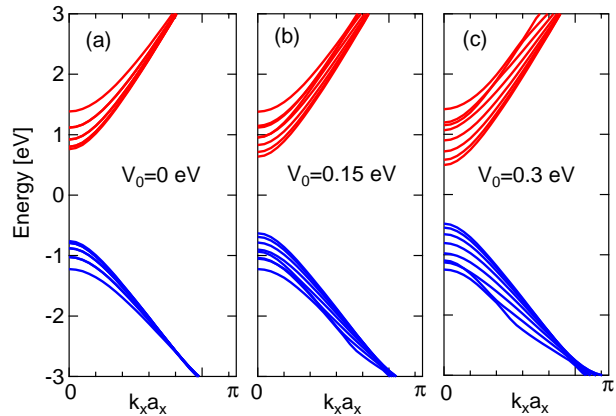}
\caption{\label{fig:BP_band} The electron band structure of the BP superlattice with $Q=20$ for (a) $V_0=0$, (b) 0.15, and (c) 0.3 eV. 16 dispersion curves around the band edges are shown.}
\end{figure}

\begin{figure}[t]
\center
\includegraphics[scale=0.45,clip]{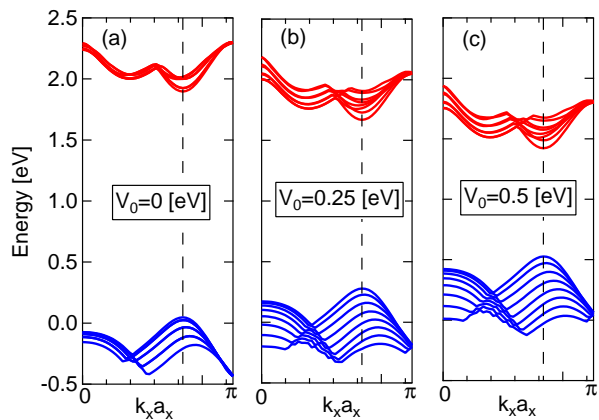}
\caption{\label{fig:SL1} The electron band structure of the MoS$_2$ superlattice with $Q=20$ for (a) $V_0=0$, (b) 0.25, and (c) 0.5 eV. 16 dispersion curves around the band edges are shown. The VB top and CB bottom are located at $k_xa_x = 2\pi /3$ (dashed line).}
\end{figure}

\subsection{Black Phosphorene}
\label{sec:BP}
First, we study the electronic structure of the BP superlattice. We use the TB model developed by Rudenko and Katsnelson.\cite{rudenko} They showed that the BP has a direct band gap at the $\Gamma$ point, where the VB top and CB bottom consist of a mixture of the $s$, $p_x$, and $p_z$ orbitals and have different orbital characters. The TB Hamiltonian of the BP superlattice is given by 
\begin{equation}
 {\cal H}_{\rm BP} = 
 \sum_{s} \left( \epsilon_s + v_s \right) c_{s}^{\dagger} c_s
 +
 \sum_{s\ne s'} t_{ss'} c_{s'}^{\dagger} c_s,
 \label{eq:BP}
\end{equation}
where the first and second terms in the right hand side are the on-site potential and kinetic energies. $c_{s}$ and $c_{s}^{\dagger}$ are the electron destruction and creation operators at the $s$th atom site, respectively. $t_{ss'}$ is the hopping integral between the sites $s$ and $s'$. These are given by $t_{14} = -1.220$, $t_{12} = 3.665$, $t_{18} = -0.205$, $t_{13} = -0.105$, and $t_{25} = -0.055$ in units of eV [see also Fig.~\ref{fig:network}(a) and (c) for the atom positions].\cite{rudenko} The 1DSL potential, $v_s$ defined by Eq.~(\ref{eq:SLpot}), is added to the TB Hamiltonian. Figures \ref{fig:BP_band}(a)-(c) show 16 dispersion curves around the band edge for $V_0 =$0, 0.15, and 0.3 eV, respectively. The zero energy is located at the middle of the CB bottom and VB top at the $\Gamma$ point by adding the on-site potential of $\epsilon_s = 0.42$ eV. When $V_0$ is increased from 0 to 0.3 eV, the band gap decreases from 1.52 to 0.97 eV. This arises from the charge redistribution within the 1DSL potential period, which will be explained in Secs.~\ref{sec:gap_v0} and \ref{sec:2DSN}. 

\begin{figure}[t]
\center
\includegraphics[scale=0.5,clip]{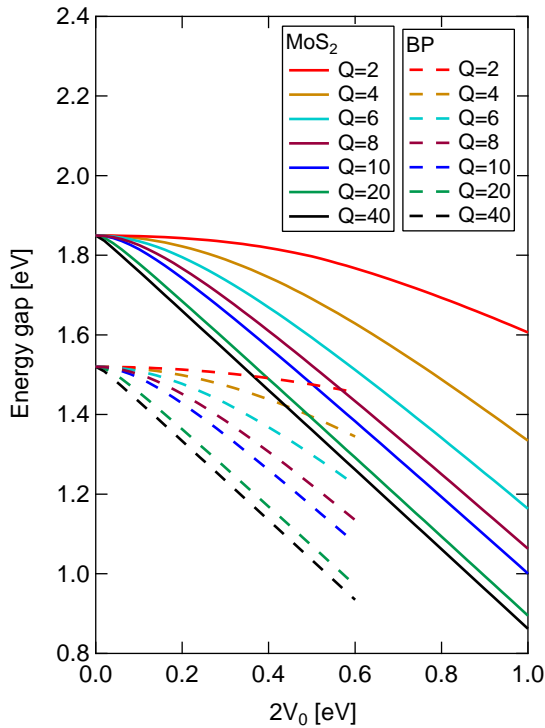}
\caption{\label{fig:Gap_M} The $2V_0$-dependence of the energy gap of MoS$_2$ (solid) and BP (dashed) superlattices for several $Q$s. }
\end{figure}

\subsection{Transition metal dichalcogenides}
\label{sec:TMD}
Next, we compute the band gap of $MX_2$ superlattice. We use the three-band TB model developed by Liu {\it et al}.\cite{liu} This model is constructed by considering the $d$-$d$ hoppings between $M$-$d_{z^2}$, $d_{xy}$, and $d_{x^2-y^2}$ orbitals, where $d_{z^2}$ ($d_{xy}$, $d_{x^2-y^2}$) is the basis of the irreducible representation of $A'_{1}$ ($E'$) for the point group $D_{3h}$. The pristine $MX_2$ exhibits a direct band gap at K and K$'$ points. The VB top mainly consists of the $d_{xy}$ and $d_{x^2-y^2}$ orbitals, while the CB bottom consists of the $d_{z^2}$ orbital. By using $\gamma$ and $\gamma'$ to denote the three atomic orbitals, the Hamiltonian of the $MX_2$ superlattice is given by
\begin{eqnarray}
 {\cal H}_{MX_2} 
 &=& 
 \sum_{s}\sum_{\gamma} \left(\epsilon_{\gamma} + v_{s} \right) c_{s,\gamma}^{\dagger} c_{s,\gamma}
 \nonumber\\
 &+&
 \sum_{s\ne s'}\sum_{\gamma,\gamma'} T_{\gamma,\gamma'}(\bm{R}^{s'} - \bm{R}^{s})
  c_{s',\gamma'}^{\dagger} c_{s,\gamma},
 \label{eq:MoS2}
\end{eqnarray}
where $c_{s,\gamma}$ and $c_{s,\gamma}^{\dagger}$ are the electron destruction and creation operators at the $s$th atom site with the energy $\epsilon_\gamma$, respectively. $T_{\gamma,\gamma'}(\bm{R})$ is the hopping integral between the orbital $\gamma$ at the site $\bm{0}$ and the orbital $\gamma'$ at the site $\bm{R}$. The summation of $s$ and $s'$ in Eq.~(\ref{eq:MoS2}) is taken up to the third nearest-neighbor (NN) sites; $\bm{R}=\bm{R}_i$ for the first NN sites, $\bm{R}=\bm{R}'_i = \bm{R}_i + \bm{R}_{i+1}$ for the second NN sites, and $\bm{R}=\bm{R}''_i = 2\bm{R}_i$ for the third NN sites, where $i=1,2,3,4,5$, and 6, with $\bm{R}_{7}=\bm{R}_1$ [see also Fig.~\ref{fig:network}(d) for the translation vectors $\bm{R}_i$]. The expressions of $3\times3$ matrix $T_{\gamma,\gamma'} (\bm{R})$ for 18 NN sites are provided in the Appendix \ref{sec:app1}. Among $MX_2$, we consider, as an example, the monolayer MoS$_2$ superlattice. We use the TB parameters of MoS$_2$ obtained by the density-functional theory calculations within the local-density approximation.\cite{liu} Then, the values of $\epsilon_\gamma$ with $\gamma = d_{z^2}$ and $\gamma = d_{xy}$, $d_{x^2-y^2}$ are set to 0.820 eV and 1.931 eV, respectively. 

Figures \ref{fig:SL1}(a)-(c) show the energy dispersion curves along the $k_x$-direction of the MoS$_2$ superlattice for various $V_0$s. 16 dispersion curves around the band edge are shown. Similar to the case of the BP superlattice, the band gap at $k_xa_x = 2\pi /3$ (vertical dotted line) drastically decreases with increasing $V_0$.

\subsection{Gap variation}
\label{sec:gap_v0}
\begin{figure}[t]
\center
\includegraphics[scale=0.45,clip]{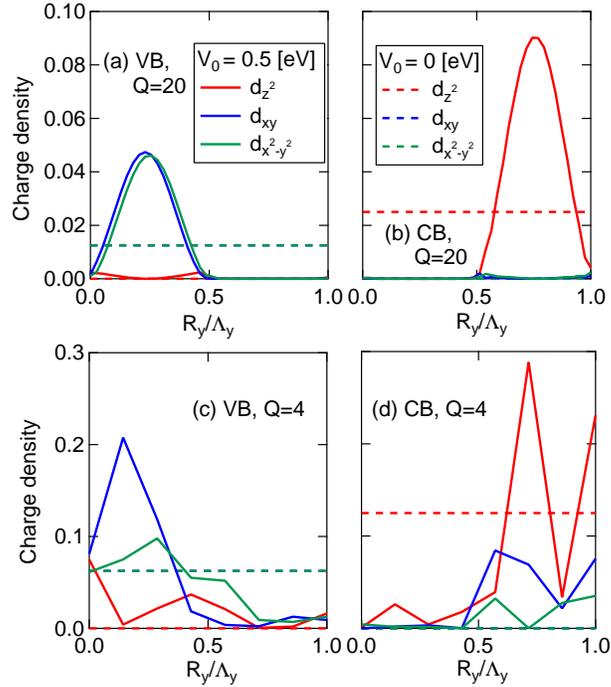}
\caption{\label{fig:charge} The $R_y$-dependence of the charge density in the MoS$_2$ superlattice (a) at the VB top for $Q=20$, (b) at the CB bottom for $Q=20$, (c) at the VB top for $Q=4$, and (d) at the CB bottom for $Q=4$, with $V_0 = 0.5$ eV (solid) and 0 eV (dashed) given by Eq.~(\ref{eq:SLpot}). The contributions from $d_{z^2}$ (red), $d_{xy}$ (blue), and $d_{x^2-y^2}$ (green) orbitals are shown. The charge distributions of $d_{xy}$ and $d_{x^2-y^2}$ completely overlap when $V_0=0$ eV.}
\end{figure}

Figure \ref{fig:Gap_M} shows the energy gap of the MoS$_2$ superlattice as a function of the barrier height $2V_0$ for various $Q$s. For comparison, the $2V_0$-dependence of the band gap of the BP superlattice is also shown. The energy gap $E_g (V_0)$ decreases monotonically with increasing $2V_0$, while the decrease in $E_g (V_0)$ is moderate for small $Q$. For larger $Q$, the band gap difference is approximately expressed by the linear relation 
\begin{equation}
 E_g(V_0) - E_g(0) \simeq - 2V_0.
 \label{eq:eg}
\end{equation}
This reflects the fact that the charges are strongly localized to the region with $v_s = V_0$ and $-V_0$ at the VB top and CB bottom, respectively. Figures \ref{fig:charge}(a) and \ref{fig:charge}(b) show the $R_y$-dependence of the charge density at the VB top and CB bottom, respectively. The contributions from $d_{z^2}$, $d_{xy}$, and $d_{x^2-y^2}$ orbitals are shown when $Q=20$ and $V_0=0.5$ eV (solid). For comparison, the case of $V_0 = 0$ eV is also shown (dashed). The charge density at the VB top consists of $d_{xy}$ and $d_{x^2-y^2}$ and is distributed around $R_y/\Lambda_y \le 1/2$, that is, the region with $v_s = V_0$ [Fig.~\ref{fig:charge}(a)], while that at the CB bottom consists of $d_{z^2}$ dominantly and is distributed around $R_y/\Lambda_y > 1/2$, that is, the region with $v_s = - V_0$ [Fig.~\ref{fig:charge}(b)]. In such a charge distribution, the eigenenergy would be linearly proportional to the on-site potential energy, resulting in the relation of Eq.~(\ref{eq:eg}). Figures \ref{fig:charge}(c) and \ref{fig:charge}(d) show the $R_y$-dependence of the charge density of the three orbitals at the VB top and CB bottom, respectively, for $Q=4$. For such a small $Q$, the charge density is finite even around $R_y/\Lambda_y > 1/2$ ($R_y/\Lambda_y \le 1/2$) at the VB top (the CB bottom). In this case, the magnitude of the upward and downward shifts of the bands is small for the VB and CB, respectively, yielding the slight decrease in the band gap.

It should be noted that the semiconductor-metal transition would be observed when $E_g(0) = 2V_0$ is satisfied in Eq.~(\ref{eq:eg}) for large $Q$s. For example, $V_0 \simeq 0.9$ and 0.75 eV is needed to observe the transition in MoS$_2$ and BP, respectively.

\section{Discussion}
\subsection{Origin of the band gap decrease}
\label{sec:2DSN}
In Sec.~\ref{sec:2DMS}, it has been shown that the 1DSL potential reduces the magnitude of the direct band gap. We attribute such a reduction to the different orbital characters between the VB top and CB bottom. As mentioned in Secs.~\ref{sec:BP} and \ref{sec:TMD}, this condition is satisfied in BP and $MX_2$. To show how the decrease in the direct band gap is explained in terms of the orbital characters at the band edges, we consider, as a simple example, a 2D square net that consists of the atoms shown in Fig.~\ref{fig:network}(e). There is an atom in the unit cell whose size is $a_x \times a_y$. Each atom has an atomic energy of $\epsilon_\gamma$. We assume that there are two energy levels $\gamma=a$ and $b$ satisfying the relation $\epsilon_a< \epsilon_b$. The $s$th atom position is denoted by $\bm{R}^s = (R_{x}^{s},R_{y}^{s}) = (na_x,ma_y)$ with integers $n$ and $m$. The different potentials are added to study the superlattice potential effect: $v_s = V_0 \ (>0)$ and $-V_0$ for $m = 2l$ and $m=2l+1$ with an integer $l$, respectively. The real-space TB Hamiltonian for the 2D square net is given by
\begin{eqnarray}
 {\cal H}_{\rm SQ} 
 = \sum_{s}\sum_{\gamma=a}^{b} \left( \epsilon_\gamma + v_s \right)
 c_{s,\gamma}^{\dagger} c_{s,\gamma}
 + \sum_{s\ne s'} \sum_{\gamma=a}^{b} t_\gamma
 c_{s',\gamma}^{\dagger}  c_{s,\gamma},
 \nonumber\\
\end{eqnarray}
where $t_\gamma$ is the electron hopping integrals between the energy levels $\epsilon_\gamma$ at the nearest-neighbor sites. The electron hopping between the energy levels $\gamma=a$ and $\gamma = b$ is assumed to be negligible. By imposing the periodic boundary condition to form the energy bands, the TB Hamiltonian in a reciprocal-space becomes a $4\times4$ matrix:
\begin{eqnarray}
 \tilde{{\cal H}}_{\rm SQ}(\bm{k}) &=& 
 \begin{pmatrix}
  h_{a}^{+} & 2t_a \cos \beta & 0 & 0 \\
 2t_a \cos \beta &  h_{a}^{-} &  0 &  0 \\
  0 & 0 & h_{b}^{+} & 2t_b \cos \beta \\
  0 &  0  & 2t_b \cos \beta &  h_{b}^{-}
 \end{pmatrix},
 \nonumber\\
 \label{eq:TB}
\end{eqnarray}
where, for $\gamma=a$ and $b$,
\begin{eqnarray}
h_{\gamma}^{+} &=& \epsilon_\gamma + V_0 +2t_\gamma \cos \alpha,
\nonumber\\
  h_{\gamma}^{-} &=& \epsilon_\gamma - V_0 +2t_\gamma \cos \alpha,
\end{eqnarray}
with $\alpha = k_x a_x$ and $\beta = k_y a_y$ being the wavevector $\bm{k} = (k_x,k_y)$. The energy eigenvalues are given by
\begin{eqnarray}
 E_{\gamma}^{\pm} (\bm{k}) &=& \epsilon_\gamma +2t_\gamma \cos \alpha \pm \sqrt{V_{0}^{2} + 4t_{\gamma}^{2}\cos^2 \beta}.
\end{eqnarray}
Below, we use the indexes ($\gamma,\pm$) to denote the energy band. We assume $t_a > 0$ and $t_b<0$, since we study the band structure with semiconducting properties. Then, both the energy maximum of the band $(a,+)$ and the energy minimum of the band $(b,-)$ are located at the $\Gamma$ point. The energy difference $\Delta E$ is explicitly given by
\begin{eqnarray}
 \Delta E 
 &=& E_{b}^{-} (\bm{k}=0) - E_{a}^{+} (\bm{k}=0)
 \nonumber\\
  &=& \epsilon_b - \epsilon_a+ 2(t_b - t_a) 
  - \sqrt{V_{0}^{2} + 4t_{b}^{2}} - \sqrt{V_{0}^{2} + 4t_{a}^{2}}.
  \nonumber\\
  \label{eq:deltaE}
\end{eqnarray}
When $\Delta E >0$, the bands ($a,+$) and ($b,-$) serve as the VB and CB, respectively. From Eq.~(\ref{eq:deltaE}), it is clear that the energy gap $\Delta E$ decreased with increasing $V_0$. This is because the charges are redistributed to obtain the energy gain. The eigenvectors of the VB top [i.e., at the $\Gamma$ point in the ($a,+$) band] and the CB bottom [i.e., at the $\Gamma$ point in the ($b,-$) band] are respectively given by
\begin{eqnarray}
 \begin{pmatrix}
  C_{a}^{+} \\
  C_{a}^{-} \\
  C_{b}^{+} \\
  C_{b}^{-}
 \end{pmatrix} 
 = \cfrac{2t_{a}}{\sqrt{w_{a,-}^{2} + 4t_{a}^{2}}}
  \begin{pmatrix}
  1 \\
 w_{a,-}/(2t_a) \\
  0 \\
  0
 \end{pmatrix}
\end{eqnarray} 
and 
\begin{eqnarray}
 \begin{pmatrix}
  C_{a}^{+} \\
  C_{a}^{-} \\
  C_{b}^{+} \\
  C_{b}^{-}
 \end{pmatrix} 
 = \cfrac{2t_{b}}{\sqrt{w_{b,+}^{2} + 4t_{b}^{2}}}
  \begin{pmatrix}
  0 \\
  0 \\
   1 \\
  - w_{b,+}/(2t_b) 
 \end{pmatrix},
\end{eqnarray} 
where $w_{\gamma,p} = \sqrt{V_{0}^{2} + 4t_{\gamma}^2} +p V_0$ with $p=\pm$ and $\gamma = a,b$. $C_{\gamma}^{+}$ and $C_{\gamma}^{-}$ are the probability amplitude of the orbital $\gamma$ at the sites $m=2l$ and $m=2l+1$, respectively. Since $w_{a,-}/(2t_a) <1$ and $- w_{b,+}/(2t_b) >1$ for finite $V_0$, the charges at the VB top and CB bottom are mainly distributed at the sites $m=2l$ ($v_s = V_0$) and $m=2l+1$ ($v_s=-V_0$), respectively. As a result, the ($a,+$) and ($b,-$) bands shift to higher and lower energies, respectively, which leads to the decrease in the band gap. Thus, we determined the relationship between the band gap and the orbital difference at the band edge. Although the 2D square net model above is quite simple, it captures the main physics behind the band gap reduction observed in Figs.~\ref{fig:BP_band} and \ref{fig:SL1}.

It is possible to construct a TB model that includes more than two atoms in a unit cell, for example, $v_s = V_0$ for $m=4l$ and $4l+1$, and $v_s= - V_0$ for $m=4l+2$ and $4l+3$ with an integer $l$. In such a case, the magnitude of the band gap decreases more significantly because more charges are redistributed within the unit cell. This is also consistent with the observation in Fig.~\ref{fig:Gap_M}.

\begin{figure}[t]
\center
\includegraphics[scale=0.45,clip]{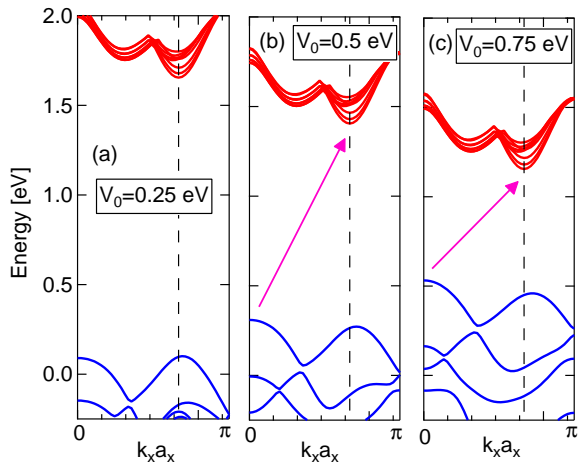}
\caption{\label{fig:indirect} The electron band structure of the MoS$_2$ under a 1DSL potential given by Eq.~(\ref{eq:SLpot2}) with $Q=20$, calculated with (a) $V_0=0.25$, (b) 0.5, and (c) 0.75 eV. }
\end{figure}

\subsection{Direct-to-indirect gap transition}
\label{sec:app2}
While we have also studied other types of 1DSL potentials, such as a cosine-type potential, where the net total potential per unit cell vanishes, the results in this work are qualitatively the same. When the net total potential per unit cell is finite, the system exhibits an indirect band gap. To show the latter, we study the MoS$_2$ under a 1DSL potential with {\it unequal} barrier width. The 1DSL potential used is given by 
\begin{equation}
 v_{s} = 
 \begin{cases}
  V_0 & {\rm for} \ \ 0\le R^{s}_{y} < \dfrac{\Lambda_y}{10} \\
  -V_0 & {\rm for} \ \ \dfrac{\Lambda_y}{10}  \le R^{s}_{y} < \Lambda_y.
 \end{cases}
 \label{eq:SLpot2}
\end{equation}
Figure \ref{fig:indirect} shows the band structure of the MoS$_2$ under a 1DSL potential given by Eq.~(\ref{eq:SLpot2}) with $\Lambda_y = 20a_y$ for various $V_0$s. As $V_0$ increases, the band gap decreases, similar to the cases of the 1DSL potential with an equal barrier width shown in Fig.~\ref{fig:SL1}. Interestingly, when $V_0 \ge 0.5$ eV, the energy of the VB top at $k_x a_x=0$ overcomes that at $k_x a_x=2\pi/3$, yielding the indirect band gap. This originates from the different orbital characters of the VB top between $k_xa_x = 0$ and $k_xa_x = 2\pi /3$ in the pristine MoS$_2$, where the former mainly consists of a Mo $d_{z^2}$ orbital, while the latter consists of Mo $d_{xy}$ and $d_{x^2-y^2}$ orbitals.\cite{liu} In addition, since the charge of the CB bottom at $k_xa_x = 2\pi /3$ is distributed around the region with $v_{s} = -V_0$, as shown in Fig.~\ref{fig:charge}(b), the charge of the VB top at $k_x a_x=0$ tends to be, in turn, distributed around the region with $v_{s} = +V_0$ (i.e., $R^{s}_{y} < \Lambda_y/10$). This leads to an increase in the energy of the VB top at $k_xa_x = 0$, compared with that at $k_xa_x = 2\pi /3$.

\section{Conclusions}
Through the TB calculations for the BP and $MX_2$ superlattices, we have demonstrated that the presence of a 1DSL potential yields a decrease in the band gap of 2D materials. An analytical investigation shows that this also holds if a pristine 2D material has (i) a direct-band gap and (ii) different orbital characters between the VB top and the CB bottom. It has also been found that the band gap experiences a direct-to-indirect gap transition when a 1DSL potential with an unequal barrier width is applied. We expect that various 2D material superlattices will be created in future experiments, as the graphene superlattices are fabricated by several experiments.\cite{dubey,bai,lin,drienovsky}

\begin{acknowledgments}
SO would like to thank M. Aoki for fruitful discussions and G.-B. Liu and D. Xiao for providing useful information about the TB model of $MX_2$. This study was supported by a Grant-in-Aid for Young Scientists B (No. 15K17435) from JSPS.
\end{acknowledgments}

\appendix
\section{Hopping integrals}
\label{sec:app1}
The matrix elements $T_{\gamma,\gamma'}(\bm{R})$ in Eq.~(\ref{eq:MoS2}) can be expressed by 17 parameters.\cite{liu} By using the notation $(d_{z^2}, d_{xy}, d_{x^2 -y^2}) = (1,2,3)$, the 17 hopping parameters for the first NN sites are explicitly given as:
\begin{widetext}
\begin{eqnarray*}
 t_0=T_{1,1}(\bm{R}_1), \ 
 t_1=T_{1,2}(\bm{R}_1), \ 
 t_2=T_{1,3}(\bm{R}_1), \
 t_{11}=T_{2,2}(\bm{R}_1), \
 t_{12}=T_{2,3}(\bm{R}_1), \ 
 t_{22}=T_{3,3}(\bm{R}_1),
 \end{eqnarray*}
for the second NN sites:
\begin{eqnarray*}
 r_0=T_{1,1}(\bm{R}'_1), \ 
 r_1=T_{1,2}(\bm{R}'_1), \
 r_2=T_{1,2}(\bm{R}'_4), \
 r_{11}=T_{2,2}(\bm{R}'_1), \
 r_{12}=T_{2,3}(\bm{R}'_1), \
 \end{eqnarray*}
and for the third NN sites:
\begin{eqnarray*}
  u_0=T_{1,1}(\bm{R}''_1),  \
 u_1=T_{1,2}(\bm{R}''_1),  \
 u_2=T_{1,3}(\bm{R}''_1), \
 u_{11}=T_{2,2}(\bm{R}''_1),  \ 
 u_{12}=T_{2,3}(\bm{R}''_1),  \
 u_{22}=T_{3,3}(\bm{R}''_1).
\end{eqnarray*}
The other matrix elements are obtained with the aid of the symmetry property. Those are explicitly written as, for the first nearest-neighbor (NN) sites $\bm{R}_i$, 
\begin{equation}
 T (\bm{R}_1) =
 \begin{pmatrix}
 t_0 & t_1 & t_2 \\
 -t_1 & t_{11} & t_{12} \\
 t_2 & - t_{12} & t_{22}
 \end{pmatrix}, \ \ 
 T (\bm{R}_2) = 
 \begin{pmatrix}
 t_0 & 2c t_1 - 2dt_2 & - 2dt_1 -2 ct_2 \\
 - 2c t_1 -2 dt_2 & ct_{11} + 3ct_{22} & -dt_{11} - t_{12} + dt_{22}\\
 2dt_1 - 2ct_2 & -dt_{11} + t_{12} + dt_{22} & 3ct_{11} + ct_{22}
 \end{pmatrix},
 \end{equation}
 \begin{equation}
 T (\bm{R}_3) = 
 \begin{pmatrix}
 t_0 & - 2c t_1 + 2dt_2 & - 2dt_1 -2 ct_2 \\
 2c t_1 + 2 dt_2 & ct_{11} + 3ct_{22} & dt_{11} + t_{12} - dt_{22}\\
 2 dt_1 -2 ct_2 & dt_{11} - t_{12} - dt_{22} & 3ct_{11} + ct_{22}
 \end{pmatrix}, \ \ 
   T (\bm{R}_4) =
 \begin{pmatrix}
 t_0 & - t_1 & t_2 \\
 t_1 & t_{11} & - t_{12} \\
 t_2 & t_{12} & t_{22}
 \end{pmatrix},
\end{equation}
 \begin{equation}
 T (\bm{R}_5) = 
 \begin{pmatrix}
 t_0 & - 2c t_1 - 2dt_2 &  2dt_1 -2 ct_2 \\
 2c t_1 - 2dt_2 & ct_{11} + 3ct_{22} & - dt_{11} + t_{12} + dt_{22}\\
 - 2dt_1 -2 ct_2 & - dt_{11} - t_{12} + dt_{22} & 3ct_{11} + ct_{22}
 \end{pmatrix},
\end{equation}
 \begin{equation}
 T (\bm{R}_6) = 
 \begin{pmatrix}
 t_0 & 2c t_1 + 2dt_2 &  2dt_1 -2 ct_2 \\
 - 2c t_1 + 2dt_2 & ct_{11} + 3ct_{22} & dt_{11} - t_{12} - dt_{22}\\
 - 2dt_1 -2 ct_2 &  dt_{11} + t_{12} - dt_{22} & 3ct_{11} + ct_{22}
 \end{pmatrix},
\end{equation}
and, for the next NN site $\bm{R}'_i = \bm{R}_i + \bm{R}_{i+1}$ with $\bm{R}_{7}=\bm{R}_1$:
\begin{equation}
 T (\bm{R}'_1) = 
 \begin{pmatrix}
 r_0 & r_1 & -e r_1 \\
 r_2 & r_{11} & r_{12} \\
 - e r_2 & r_{12} & r_{11} + 2e r_{12}
 \end{pmatrix}, \ \ 
  T (\bm{R}'_2) = 
 \begin{pmatrix}
 r_0 & 0 & 2e r_1 \\
 0 & r_{11} + r_{12}/e & 0 \\
 2e r_1 &0 & r_{11} - e r_{12}
 \end{pmatrix}, 
\end{equation}
\begin{equation}
 T (\bm{R}'_3) = 
 \begin{pmatrix}
 r_0 & - r_1 & -e r_1 \\
 - r_2 & r_{11} & - r_{12} \\
 - e r_2 & - r_{12} & r_{11} + 2e r_{12}
 \end{pmatrix}, \ \ 
  T (\bm{R}'_4) = 
 \begin{pmatrix}
 r_0 & r_2 & - e r_2 \\
 r_1 & r_{11} & r_{12} \\
 - e r_1 & r_{12} & r_{11} + 2 e r_{12}
 \end{pmatrix}, 
\end{equation}
\begin{equation}
 T (\bm{R}'_5) = 
 \begin{pmatrix}
 r_0 & 0 &  2 e r_1 \\
 0 & r_{11} + r_{12}/e & 0 \\
 2 e r_2 & 0 & r_{11} - e r_{12}
 \end{pmatrix}, \ \ 
  T (\bm{R}'_6) = 
 \begin{pmatrix}
 r_0 & - r_2 & - e r_2 \\
 - r_1 & r_{11} & - r_{12} \\
 - e r_1 & - r_{12} & r_{11} + 2 e r_{12}
 \end{pmatrix},
\end{equation}
\end{widetext}
with $c=1/4, d=\sqrt{3}/4$ and $e= 1/\sqrt{3}$. The expressions for the third NN sites $\bm{R}''_i = 2\bm{R}_i$ can be obtained by replacing $(t_0,t_1,t_2,t_{11},t_{12},t_{22})$ in the expressions for the first NN sites $\bm{R}_i$ with $(u_0,u_1,u_2,u_{11},u_{12},u_{22})$, respectively.


\end{document}